\documentclass{article}

\usepackage{arxiv}

\usepackage[utf8]{inputenc} 
\usepackage[T1]{fontenc}    
\usepackage{hyperref}       
\usepackage{url}            
\usepackage{booktabs}       
\usepackage{amsfonts}       
\usepackage{nicefrac}       
\usepackage{microtype}      
\usepackage{lipsum}		
\usepackage{graphicx}
\usepackage{natbib}
\usepackage{doi}
\usepackage{siunitx}
\usepackage{amsmath,amssymb,amsfonts}
\usepackage{mathtools} 

\usepackage{amssymb}
\usepackage{pifont}
\usepackage{float}
\usepackage{caption}
\usepackage{subcaption}
\usepackage{comment}
\usepackage{mathtools}

\title{Hardware-Algorithm Co-design Enabling Processing-in-Pixel-in-Memory (P$^2$M) for Neuromorphic Vision Sensors}

\author{Md Abdullah-Al Kaiser \\
	University of Southern California\\
	Los Angeles, CA 90089 \\
	\texttt{mdabdull@usc.edu} \\
 	\And
	Akhilesh R. Jaiswal \\
	University of Wisconsin-Madison\\
	Madison, WI 53706\\
	\texttt{akhilesh.jaiswal@wisc.edu} \\
}

\renewcommand{\shorttitle}{Neuromorphic P$^2$M Co-design}

\begin{document}
\maketitle

\begin{abstract}
The high volume of data transmission between the edge sensor and the cloud processor leads to energy and throughput bottlenecks for resource-constrained edge devices focused on computer vision. Hence, researchers are investigating different approaches (e.g., near-sensor processing, in-sensor processing, in-pixel processing) by executing computations closer to the sensor to reduce the transmission bandwidth. Specifically, in-pixel processing for neuromorphic vision sensors (e.g., dynamic vision sensors (DVS)) involves incorporating asynchronous multiply-accumulate (MAC) operations within the pixel array, resulting in improved energy efficiency. In a CMOS implementation, low overhead energy-efficient analog MAC accumulates charges on a passive capacitor; however, the capacitor's limited charge retention time affects the algorithmic integration time choices, impacting the algorithmic accuracy, bandwidth, energy, and training efficiency. Consequently, this results in a design trade-off on the hardware aspect- creating a need for a low-leakage compute unit while maintaining the area and energy benefits. In this work, we present a holistic analysis of the hardware-algorithm co-design trade-off based on the limited integration time posed by the hardware and techniques to improve the leakage performance of the in-pixel analog MAC operations. 
\end{abstract}

\keywords{Neuromorphic Vision Sensors, Convolution, In-pixel Processing,  Hardware-Algorithm Co-design, DVS128-Gesture.}

\maketitle

\section{Introduction}
Energy inefficiency and throughput bottlenecks in edge devices equipped with computer vision systems stem from the high volume of data transmission due to the physical segregation of the sensing hardware and computing platform \cite{system_bottleneck}. To reduce the data transmission bandwidth, researchers are exploring near-sensor, in-sensor, and in-pixel processing approaches by bringing the computations closer to the sensor \cite{near_sensor_3D_sony, sleepspotter, aps_p2m, datta2022p2mdetrack, datta2023icassp}. Though most of the research works focus on conventional frame-based CMOS image sensors (CIS), many researchers are exploring the use of event-driven neuromorphic vision sensors \cite{DVS_ref1, DVS_ref2} for different neural network applications (e.g., autonomous driving \cite{DVS_auto_driving}, pose re-localization \cite{DVS_pose}, steering angle prediction \cite{DVS_steering}, etc.) due to their energy, latency, and throughput advantages over traditional CMOS imagers. Moreover, neuromorphic vision sensors can adapt to the illumination level of the scene due to their high dynamic range compared to the conventional CIS. All these advantages collectively motivate a paradigm shift toward incorporating neuromorphic vision sensors in vision-based applications. 

Neuromorphic vision sensors often utilize spiking convolutional neural networks (CNN) to process asynchronous input events. Conventionally, the complete dataset duration is divided into multiple fixed integration times. The number of input spikes is accumulated in each integration time for each pixel, creating multi-bit inputs to the spiking CNN. Hence, the first layer of the spiking CNN comprises digital multi-bit MAC operations, in contrast to the subsequent spiking CNN layers that consist of energy-efficient accumulators. To improve the energy efficiency of such systems, an in-pixel asynchronous spatio-temporal analog convolution approach has been proposed \cite{neuromorphic_p2m} that reports $\sim2\times$ backend-processing energy improvement for the IBM DVS128-Gesture dataset. This work utilizes transistors as weights and passive capacitors as accumulators to compute the analog convolution energy-efficiently. As CMOS circuits are inherently prone to leakage, achieving a long integration time or membrane potential of the spiking CNN model utilizing a low-overhead analog circuit is challenging. Increasing the integration time demands significant circuit overhead (e.g., large capacitors, active feedback circuits, etc.) that can result in large area and high energy consumption. In contrast, algorithmic accuracy, bandwidth, and training efficiency depend on the integration time. Hence, it becomes a co-design trade-off between the hardware to design low-leakage compute units to support long integration time and the algorithm to achieve accuracy close to the baseline with a shorter integration time. To this end, we present a holistic analysis of the hardware-algorithm co-design trade-off based on the choice of integration time and leaky behavior of the neuromorphic P$^2$M compute circuits. 

\section{Enablers for Neuromorphic P$^2$M }
Two key enablers of our neuromorphic P$^2$M are 3D integration technology and the advancements in analog techniques for MAC operation \cite{3D_samsung, 3D_sony, Icancel}. The neuromorphic P$^2$M framework can be heterogeneously 3D integrated where the top die comprises the DVS pixels, and the bottom die consists of the MAC compute units aligned vertically to each DVS pixel per CNN filter. DVS pixels generate ON (OFF) events when the light intensity detected by the photodiode increases (decreases) by a certain threshold. The output channels (ON and OFF) of the DVS are connected via hybrid Cu-Cu bonding with the gate terminal of the transistor that represents the weight value of the filter used in the spiking CNN model. The transistor's geometry can be modulated according to the absolute weight value of the CNN filer (e.g., a large width of the transistor represents the large value of the filter's weight). The positive (negative) weight is implemented by supplying (dumping) charges to (from) a passive capacitor per CNN filter. The passive capacitor acts like an analog accumulator, accumulating the charges for a fixed integration time. As the input spikes are binary, the accumulation voltage either steps up (positive weight) or down (negative weight) by an amount, depending on the weight values. After the integration time, the accumulated voltage is compared with a certain threshold value to generate the output activation of each filter. 

\begin{figure}[!b]
\begin{center}
\includegraphics[width=0.75\linewidth]{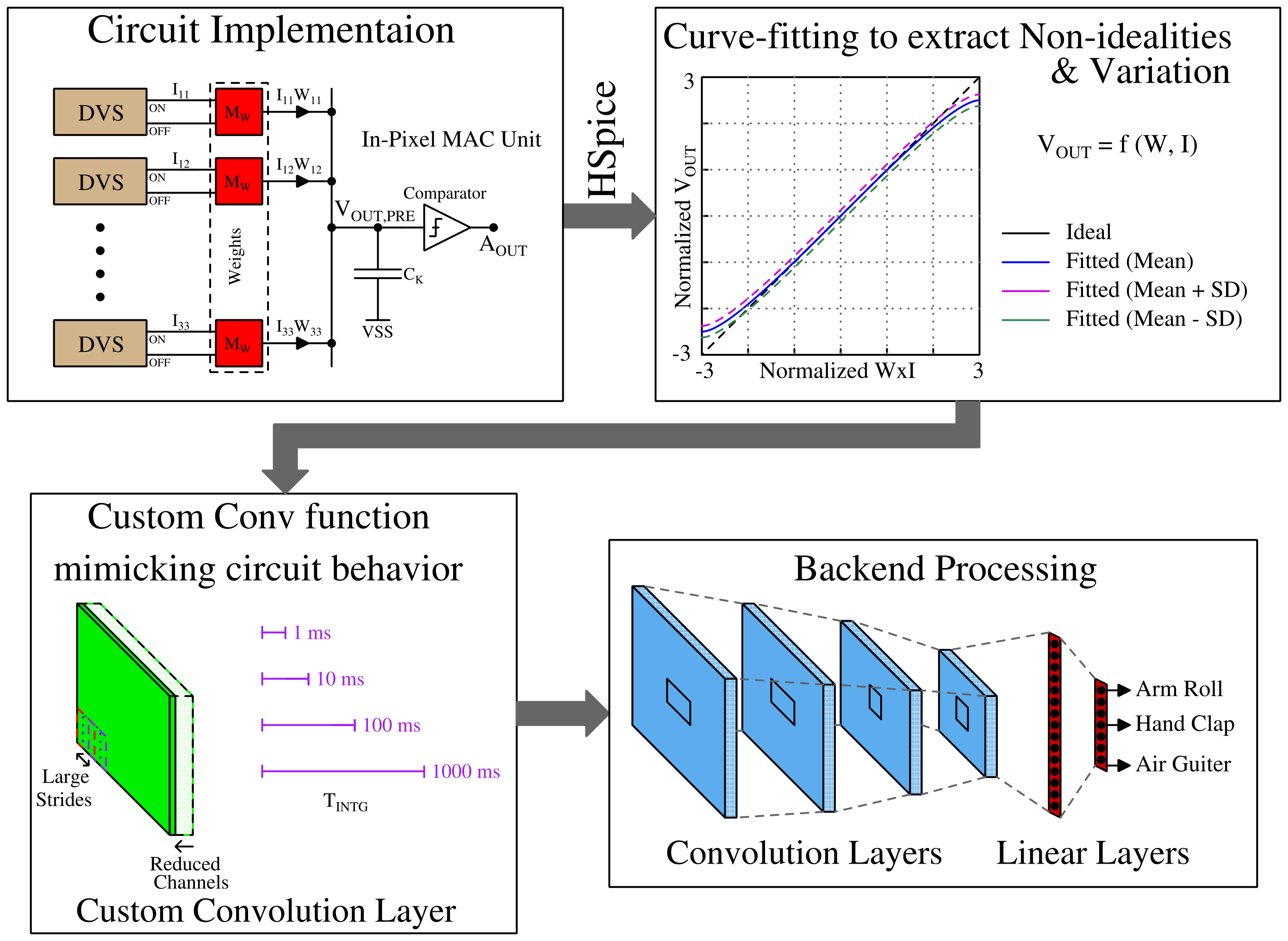}
\end{center}
\caption{Hardware-Algorithm co-design framework to enable the P$^2$M paradigm for the neuromorphic vision sensors.}
\label{fig_hardware_algorithm_codesign}
\end{figure}

Figure \ref{fig_hardware_algorithm_codesign} presents the framework for evaluating our proposed P$^2$M approach. Conventionally, the first layer of the spiking CNN comprises a linear convolution followed by a non-linear activation. In our approach, we implement the convolution utilizing voltage accumulation through inherently non-linear and appropriately sized transistors that represent the weights of the CNN filter. Consequently, any analog convolution circuit constructed with transistor devices will inherently exhibit non-ideal, non-linear characteristics. Additionally, process variations can lead to changes in the transistor's geometry that can introduce deviations in the pre-activation output (the sum of weighted multiplications) from its ideal value. Hence, we have constructed a model for our first convolution layer by employing a curve-fitting function that accounts for non-linearity, non-ideality, and process variations based on the spice simulation results using commercial GlobalFoundries PDK (GF22FDX). Furthermore, we have employed fewer channels in the first layer to maintain the DVS pixel pitch without incurring any overhead due to the analog MAC units. Readers are encouraged to check the work \cite{neuromorphic_p2m} for further insights. 

\section{Hardware-Algorithm Co-design Trade-off on Integration Time}
Leakage is a common issue in CMOS circuits, causing passive capacitors to be unable to retain accumulated charges for an extended duration. One can mitigate leakage by employing a large capacitance value; however, this approach can incur a large area overhead and reduced pixel density. As a result, the analog implementation of the hardware can limit the integration time utilized in the algorithmic framework. Typically, a long integration time is favorable for the algorithm to achieve better accuracy and for simpler training. We have evaluated the accuracy, bandwidth, backend energy, and training efficiency for various integration times considering the two large-scale popular neuromorphic benchmarking datasets (DVS128-gesture \cite{dvs-gesture}, NMNIST \cite{nmnist}). 

We use the Spikingjelly package \cite{SpikingJelly} to process the data and integrate them into a fixed time interval of 1 ms to 1000 ms based on the kernel's capacitor retention time supported by our neuromorphic P$^2$M circuit. However, it is important to note that such a short integration time results in increased timesteps (e.g., 5000 timesteps considering an integration time of 1 ms for a dataset duration of 5 s) for the neuromorphic datasets used in this study. This would significantly exacerbate the training complexity. To address this concern, we first pre-train a spiking CNN model using a long integration time, spanning seconds (small number of timesteps), without imposing P$^2$M circuit constraints. After that,  we reduce the integration time for the first spiking CNN layer to facilitate our custom P$^2$M implementation. We utilize a long integration time (in order of seconds) from the second layer, enabling the network from the second layer to process input with only a few time steps. We optimize this network starting from the second layer while keeping the first layer static because training the first layer considerably escalates memory requirements due to many timesteps. We use four convolutional layers, followed by two linear layers at the end with 512 and 10 neurons, respectively. Each convolutional layer is followed by a batch normalization layer, spiking LIF layer, and max pooling layer.

It can be observed from table \ref{tab_dvs_accuracy_results}  that a shift in integration time from 1000 ms to 1 ms can lead to an accuracy degradation of 1.65\% and 2.27\% for the NMNIST and DVS128-gesture dataset, respectively. Moreover, training efficiency depends on the number of timesteps, which is inversely related to the integration time. The normalized training time (considering 1000 ms as the baseline) becomes approximately 16.6$\times$ and $>$20$\times$ slower at the integration time of 1 ms for the NMNIST and DVS128-Gesture datasets, respectively. 

\begin{table}[!t]
\caption{Comparison of the test accuracy and training efficiency of our P$^2$M enabled spiking CNN models.}
\label{tab_dvs_accuracy_results}
\begin{center}
\scalebox{1.0}{
\begin{tabular}{l|c|c|c}
\hline
Dataset & Integration  & Accuracy  & Normalized         \\
        &  Time (ms)   &           & Training Time      \\
\hline
NMNIST  & 1         & 93.44     & 16.6$\times$          \\
        & 10        & 93.87     & 7.3$\times$           \\
        & 100       & 94.36     & 2.5$\times$           \\
        & 1000      & 95.09     & 1$\times$             \\
\hline     
DVS128- & 1         & 88.54     & ${>}$20$\times$       \\
Gesture & 10        & 88.72     & 12.0$\times$          \\
        & 100       & 89.90     & 3.8$\times$           \\
        & 1000      & 90.81     & 1$\times$             \\
\hline
\vspace{-0.2in}
\end{tabular}}
\end{center}
\end{table}

\begin{figure}[!t]
\begin{center}
\includegraphics[width=0.75\linewidth]{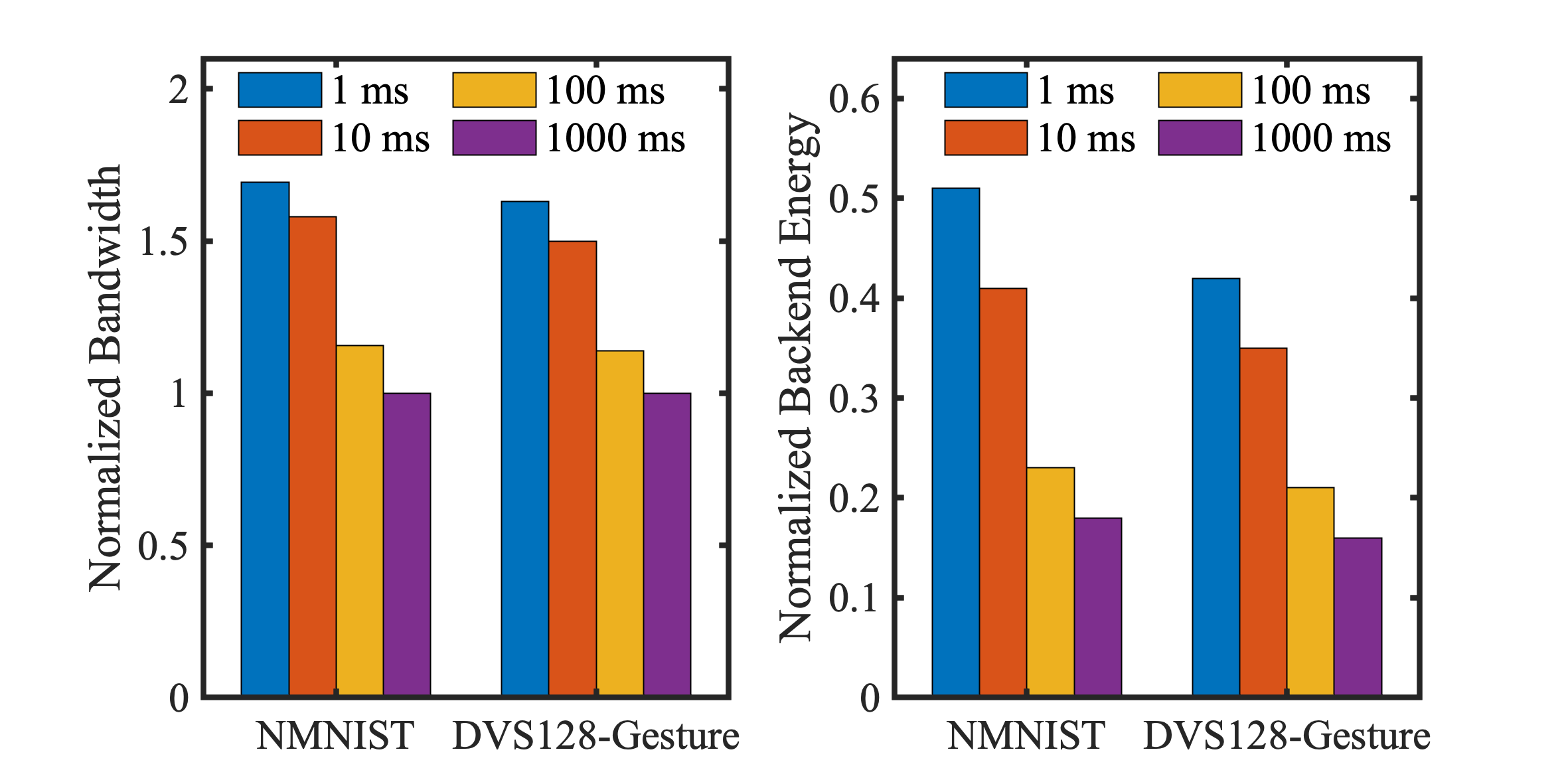}
\end{center}
\caption{Normalized bandwidth (left-subplot) and backend energy (right-subplot) for various integration times ($T_{INTG}$) considering NMNIST and DVS128-Gesture datasets.}
\vspace{-0.2in}
\label{fig_energy_bw}
\end{figure}

Figure \ref{fig_energy_bw} illustrates the normalized bandwidth and the normalized backend energy consumption across various integration times in our neuromorphic P$^2$M approach. Bandwidth has been calculated from the ratio of the average output activation spikes to input event spikes and then normalized with respect to the integration time of 1000 ms. The number of output activation spikes increases with the decrease in integration time to maintain good algorithmic accuracy. The resulting normalized bandwidth is 1.7$\times$ and 1.63$\times$ at the integration time of 1 ms for the NMNIST and DVS128-Gesture datasets, respectively. The backend compute energy consumption is normalized with respect to the conventional backend processing (i.e., digital implementation). Due to the in-pixel processing of the first spiking CNN layer in the analog domain, our neuromorphic P$^2$M approach yields at least $\sim$2$\times$ improvement in the backend compute energy in both datasets. Additionally, the advantage in compute efficiency becomes even more pronounced with longer integration times, as fewer output spikes are generated, reducing the energy needed for transmission. The backend compute energy consumption improvement varies from 2.4$\times$ (1.96$\times$) to 6.25$\times$ (5.56$\times$) for the DVS128-Gesture (NMNIST) datasets, considering the integration time ranging from 1 ms to 1000 ms. In short, a long integration time (or lowering the leakage for the accumulation capacitor in the hardware) yields low transmission bandwidth and higher backend compute efficiency. Note that the backend energy is determined by the number of spikes generated by each layer, similar to \cite{datta2021hsi, datta2021ijcnn}; more details on the method used for energy consumption estimation can be found in \cite{neuromorphic_p2m}.

\section{Low-leakage Convolution Implementation}
\begin{figure}[!b]
\centering
    \begin{subfigure}[t]{0.33\textwidth}
    \centering
    \includegraphics[width=\textwidth]{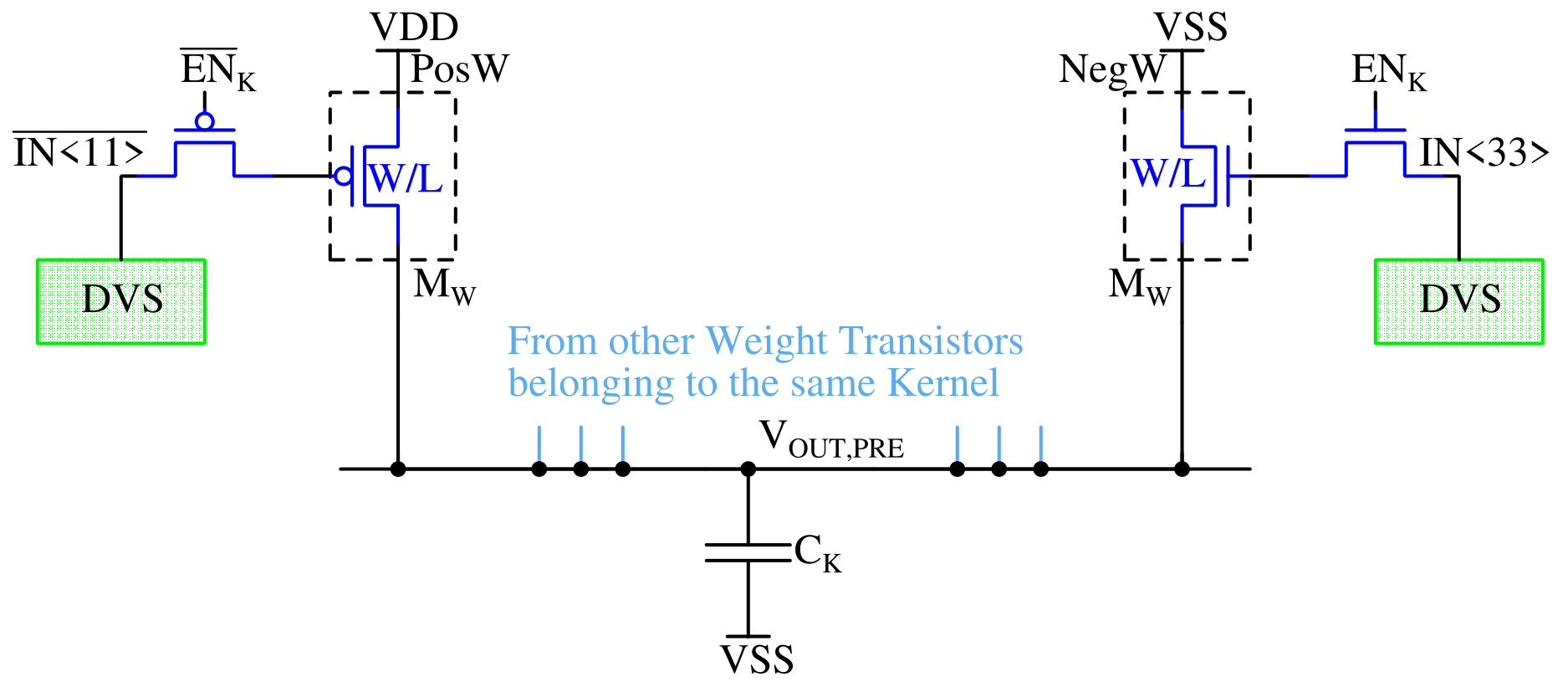}
    \caption{config-(a)}
    \label{fig_mac_basic}
    \end{subfigure}
\hfill
    \begin{subfigure}[t]{0.33\textwidth}
    \centering
    \includegraphics[width=\textwidth]{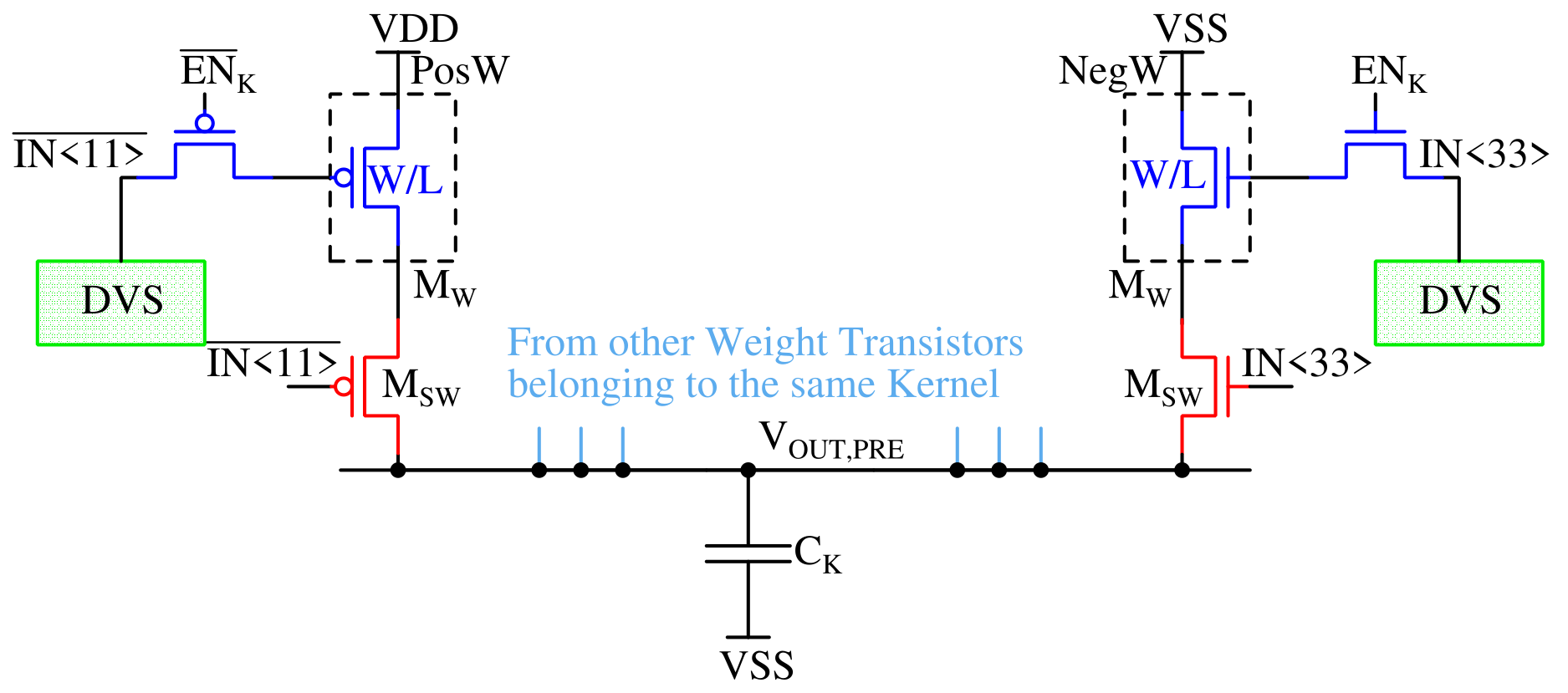}
    \caption{config-(b)}
    \label{fig_low_leakage_mac_v1}
    \end{subfigure}
\hfill
    \begin{subfigure}[t]{0.33\textwidth}
    \centering
    \includegraphics[width=\textwidth]{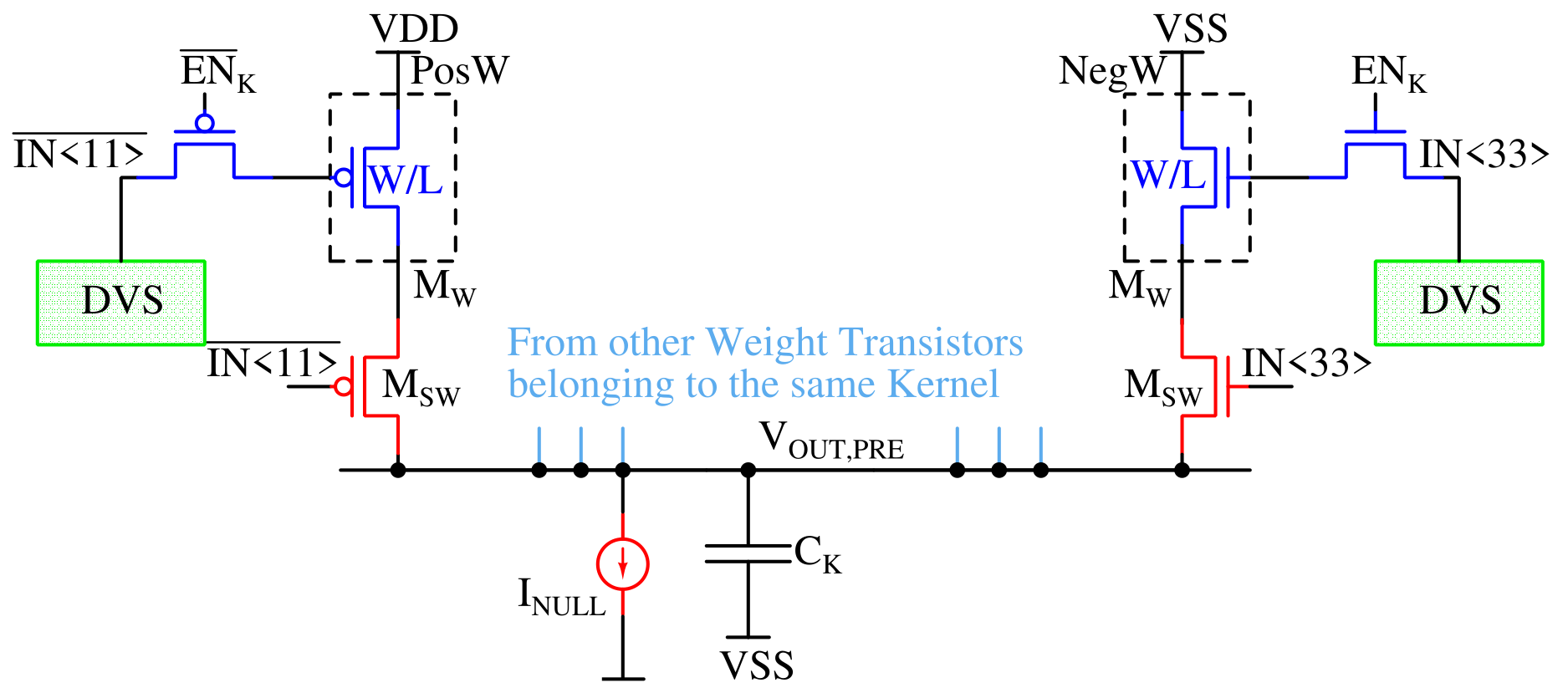}
    \caption{config-(c)}
    \label{fig_low_leakage_mac_v2}
    \end{subfigure}
\caption{Asynchronous convolution compute unit per kernel for the neuromorphic P$^2$M approach. (a) basic convolution unit consists of a multi-bit weight and accumulation capacitor, (b) the addition of a switch ($M_{SW}$) to disconnect the leakage path from the accumulation capacitor, (c) the addition of kernel-dependent nullifying current source ($I_{NULL}$) that cancels the leakage of a kernel.}
\label{fig_conv_circuit}
\end{figure}

\begin{figure*}[!t]
\begin{center}
\includegraphics[width=\linewidth]{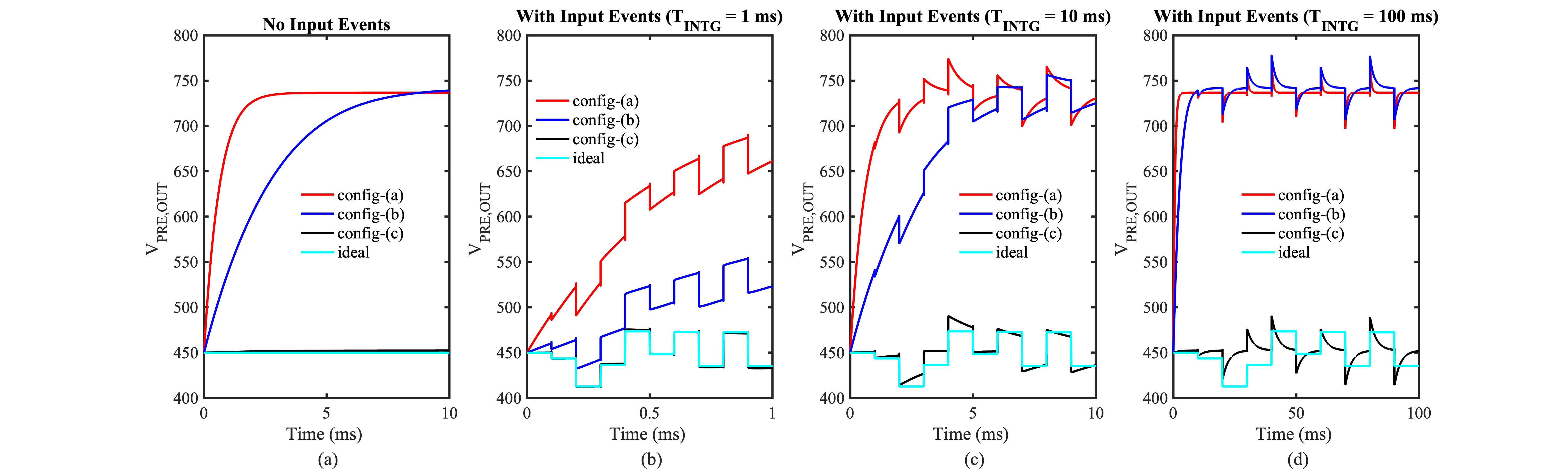}
\end{center}
\vspace{-0.2in}
\caption{Pre-activation output voltage for various integration times ($T_{INTG}$) and input events considering random kernel weights and size of 3x3. (a) With no input event spikes at $T_{INTG}$ = 10 ms, with input event spikes at (b) $T_{INTG}$ = 1 ms, (c) $T_{INTG}$ = 10 ms, and (d) $T_{INTG}$ = 100 ms.}
\vspace{-0.2in}
\label{fig_vout_leakage}
\end{figure*}

Fig. \ref{fig_conv_circuit} illustrates the MAC computation unit of our neuromorphic P$^2$M framework. Fig. \ref{fig_mac_basic} presents the baseline circuit, in which the kernel capacitor ($C_K$) accumulates the charges during the integration time, and the amount of charge depends on the weight transistor's ($M_W$) geometry. For a long integration time, the kernel capacitor will leak through the weight transistors. To reduce the leakage, we have added another switch ($M_{SW}$) between the weight transistors and kernel capacitor to disconnect the leakage path of the kernel capacitor from the weight transistors. Fig. \ref{fig_low_leakage_mac_v1} exhibits the modified version of the MAC computation unit that can yield better leakage performance than the basic unit. While including an extra switch ($M_{SW}$) helps mitigate leakage, it might not suffice for sustaining sub-millisecond integration times, especially considering that typical neuromorphic event-driven datasets often span several seconds. Consequently, we have introduced a nullifying current source ($I_{NULL}$) that provides an equivalent leakage current in the opposite direction, ensuring that the total charge on the capacitor remains unchanged (Fig. \ref{fig_low_leakage_mac_v2}). This circuit variant can reduce leakage and maintain capacitor charge for several milliseconds, making it a viable hardware choice for the neuromorphic P$^2$M approach. While improved leakage performance can be attained by employing larger capacitors or active components, this option may not be practical due to the substantial increase in area and energy demands. 

Figure \ref{fig_vout_leakage} illustrates the leakage characteristics associated with random kernel weights and input events of size 3x3. In Figure \ref{fig_vout_leakage}a, the pre-activation output voltage ($V_{OUT,PRE}$) is observed for three different circuit configurations considering no input event spikes. Notably, the sub-figure demonstrates that $V_{OUT,PRE}$ saturates in config-(a) due to leakage. The amount and direction of leakage depend on the weight values and types; positive weights are implemented using pFETs, supplying current to the kernel capacitor, hence causing the leakage toward supply, while negative weights are implemented using nFETs, which draw current from the capacitor, hence, causing the leaking toward the ground. Consequently, the combination of positive and negative weights of a CNN filter determines the direction and amount of the leakage current. Config-(b) exhibits better leakage performance than config-(a), although the capacitor still experiences significant leakage when integrating over the millisecond range. In contrast, config-(c) shows excellent leakage behavior for a 10 ms integration time, as depicted in the subfigure, as the pre-activation output voltages closely align with ideal values. 

Figures \ref{fig_vout_leakage}b, c, and d display the pre-activation output voltages for integration times of 1 ms, 10 ms, and 100 ms for the three circuit configurations. It is evident from the subfigures that configuration (c) progressively struggles to effectively mitigate leakage due to the increase in the integration time, resulting in a deviation from the ideal output voltages. While a 1 ms integration time is optimal for hardware considerations, it significantly increases the number of timesteps for the datasets spanning in the range of seconds. Table \ref{tab_dvs_accuracy_results} and Fig. \ref{fig_energy_bw} indicate that long integration times yield better algorithmic performance regarding accuracy, bandwidth, backend energy, and training efficiency. However, while offering algorithmic benefits, a 100 ms integration time may introduce significant errors that could significantly drop model accuracy. In summary, a 10 ms integration time turns out to be the most optimal choice, balancing a reasonable number of timesteps for the algorithm and acceptable leakage in the hardware. This result also highlights that neuromorphic in-pixel processing is inherently a hardware-algorithm co-design challenge that requires simultaneously optimizing both hardware and algorithm aspects, using novel low-overhead hardware techniques guided by the algorithm.

\section{Conclusion}
In this work, we have studied the hardware-algorithm co-design trade-off of the P$^2$M paradigm for the neuromorphic vision sensors based on the important parameter- integration time of the spiking CNN model. Long integration time is favorable by the algorithm due to exhibiting better accuracy, bandwidth, and energy consumption number; however, supporting a long integration time in the typical analog circuit with passive capacitors without extensive area and energy overhead is challenging. Moreover, the additional circuit complexity and the requirement for larger capacitors can offset the energy advantages of the neuromorphic P$^2$M approach and potentially impact the pixel pitch. We achieve a reasonably long integration time by utilizing two circuit techniques- isolating the leakage path from the CNN filter's capacitor and adding a nullifying current source to cancel the kernel-dependent leakage. Thus, we can support a longer integration time, hence, a reasonable number of timesteps for the algorithm to achieve close to the state-of-the-art accuracy. Notably, non-volatile memory (NVM) could play a pivotal role in extending the integration time, with the caveat of transitioning from mature foundry standard CMOS technology to emerging technology that we plan to explore in future work. 

\vspace{0.03in} 
\section{Acknowledgement:} The authors would like to thank Gourav Datta for developing the P$^2$M-constrained algorithmic framework and help in evaluating the performance matrices.

\bibliographystyle{unsrtnat}
\bibliography{references}

\end{document}